\documentclass[5p,twocolumn]{elsarticle}

\usepackage{mathptmx}
\usepackage{physics}
\usepackage{amsmath}
\usepackage{siunitx}
\usepackage[version=4]{mhchem}

\usepackage{hyperref}
\hypersetup{
    colorlinks=true,
    linkcolor=blue,
    filecolor=blue,      
    urlcolor=blue,
    citecolor=blue,
}

\usepackage{color,soul}
\usepackage{xspace}
\usepackage{xcolor}

\renewcommand{\vec}[1]{\ensuremath{\boldsymbol{\mathbf{#1}}}}

\journal{Journal of Non-Crystalline Solids: X}

\begin{document}

\begin{frontmatter}

\title{Structure Factor of a Phase Separating Binary Mixture with Natural and Forceful Interconversion of Species}

\author[1]{Thomas J. Longo\corref{cor1}}
\ead{tlongo1@umd.edu}
\author[2]{Nikolay A. Shumovskyi}
\ead{nshum@bu.edu}
\author[1,3]{Salim M. Asadov}
\author[2,4]{Sergey V. Buldryev}
\ead{buldyrev@yu.edu}
\author[1,5]{Mikhail A. Anisimov}
\ead{anisimov@umd.edu}

\cortext[cor1]{Corresponding author}

\address[1]{Institute for Physical Science and Technology, University of Maryland, College Park, MD 20742, USA}
\address[2]{Department of Physics, Boston University, Boston, MA 02215, USA}
\address[3]{Institute of Catalysis and Inorganic Chemistry, Azerbaijan National Academy of Sciences, Baku, AZ1143, AZR}
\address[4]{Department of Physics, Yeshiva University, New York, NY 10033, USA}
\address[5]{Department of Chemical and Biomolecular Engineering, University of Maryland, College Park, MD 20742, USA}

\begin{abstract}
Using a modified Cahn-Hilliard-Cook theory for spinodal decomposition in a binary mixture that exhibits both diffusion and interconversion dynamics, we derive the time-dependent structure factor for concentration  fluctuations. We compare the theory and obtain a qualitative agreement with simulations of the temporal evolution of the order parameter and structure factor in a nonequilibrium Ising/lattice-gas hybrid model in the presence of an external source of forceful interconversion. In particular, the characteristic size of the steady-state phase domain is predicted from the lower cut-off wavenumber of the amplification factor in the generalized spinodal-decomposition theory.
\end{abstract}

\begin{keyword}
Spinodal Decomposition, Phase Separation, Structure Factor, Interconversion of Species
\end{keyword}

\end{frontmatter}

\section{Introduction}
C. Austin Angell's pioneering work published 50 years ago hypothesized that the thermodynamic anomalies of liquid water could be attributed to the existence of two supramolecular states \cite{Angell_TwoState_1971}. Later, this idea was further developed to predict liquid-liquid phase separation in supercooled water below the temperature of homogeneous ice formation \cite{Angell_Amorphous_2004,Gallo_Tale_2016}, as well as in the vitreous state of various substances \cite{Sastry_Liquid_2003,Xu_Thermodynamics_2006,Bhat_Vitrification_2007,Xu_Monatomic_2009,Lascaris_Search_2014,Lascaris_Diffusivity_2015}. This phenomenon, known as ``liquid and glassy polyamorphism'' \cite{Pool_Polyamorphic_1997,Tanaka_Liquid_2020}, could be attributed to the interconversion between two alternative molecular or supramolecular states \cite{anisimov_thermodynamics_2018}. In this work, we investigate the effects of interconversion between alternative species in a binary mixture on the phase separation and phase domain growth. 

Previous studies of a hybrid Ising/lattice-gas model exhibiting both interconversion and diffusion dynamics \cite{Shum_Phase_2021} and a chiral model with interconversion of species \cite{Petsev_Effect_2021,Uralcan_Interconversion_2020} have demonstrated that interconversion dynamics breaks the symmetry of phase separation. As a result, depending on the rate of interconversion, to circumvent the energetically unfavorable formation of an interface, one phase will grow at the expense of the other, a phenomenon known as ``phase amplification'' \cite{Shum_Phase_2021,MFT_PT_2021}. However, if the alternative species are forced to interconvert due to an external source, then the formation of interfaces between species may become more favorable and the system may phase separate into steady-state microphase domains \cite{MFT_PT_2021}. Previous studies of a phase separating binary-lattice in the presence of an external interconversion force \cite{glotzer_monte_1994,glotzer_reaction-controlled_1995,Christensen_Segregation_1996} and a chiral model with dissipative intermolecular forces \cite{Uralcan_Interconversion_2020} have demonstrated steady-state microphase separation.

In this work, we consider a binary mixture that, when quenched along critical composition below the critical temperature of demixing, will phase separate via spinodal decomposition \cite{cahn_phase_1965}. Simultaneously, the alternative species may interconvert either naturally or forcefully. To characterize the formation of phase domains, we calculate the temporal evolution of the structure factor for the concentration fluctuations. We describe the time-dependent structure factor through two characteristic wavenumbers corresponding to the maximum and lower cut-off wavenumbers in the characteristic growth rate. We compare the theory to simulations of a nonequilibrium hybrid Ising/lattice-gas model exhibiting natural and forceful interconversion in addition to diffusion and obtain a qualitative agreement. 

\section{Modified Cahn-Hilliard-Cook Theory}
In this section, we generalize the Cahn-Hilliard theory of spinodal decomposition to allow for both natural and forceful interconversion of species and derive the structure factor for concentration fluctuations.

\subsection{Generalized Spinodal Decomposition}
\begin{figure}[h!]
    \centering
    \includegraphics[width=\linewidth]{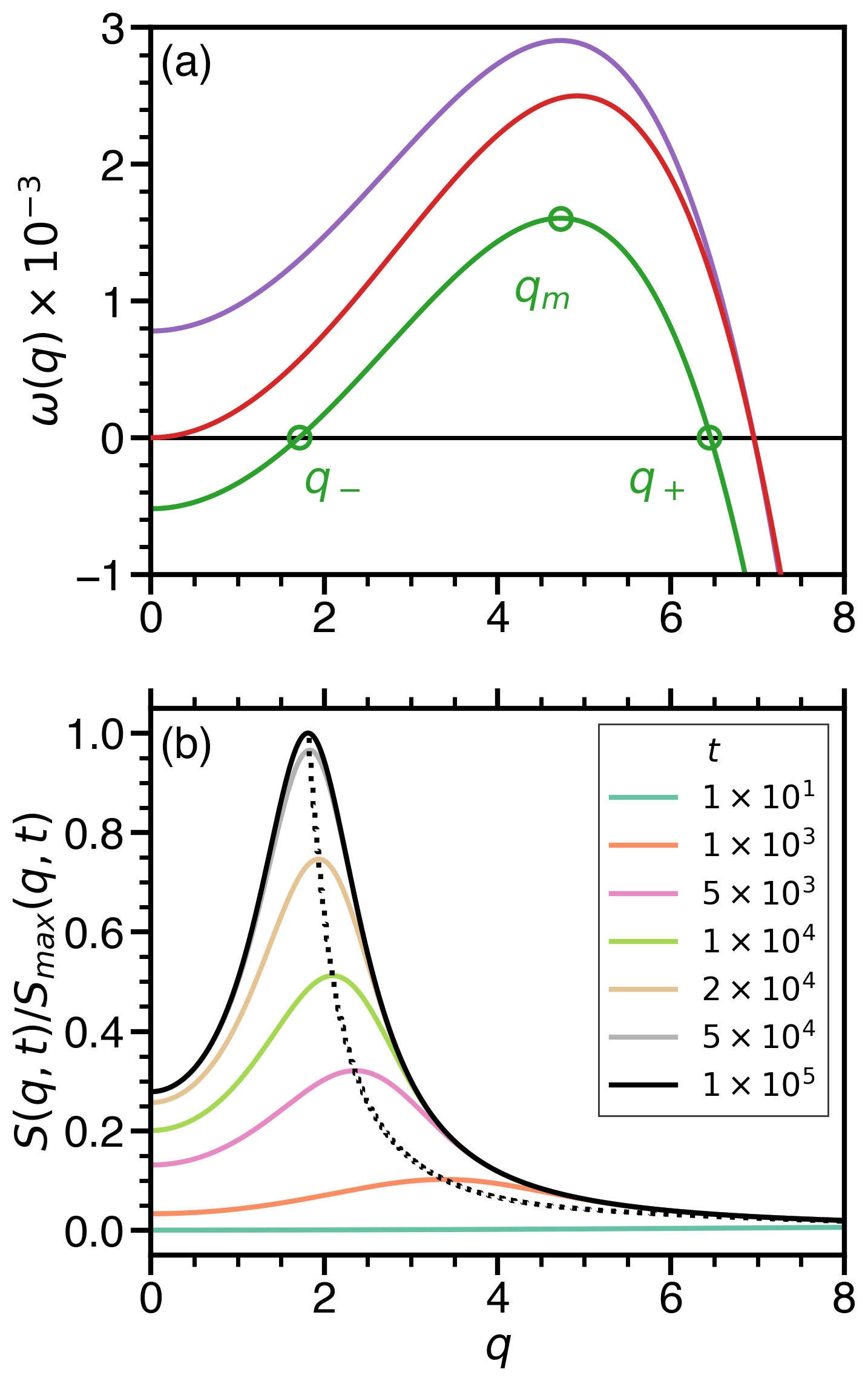}
    \caption{ a) The amplification factor, $\omega(q)$, given by Eq.~(\ref{Eqn-omega(q)}), for $\kappa = 1$ and $\Delta \hat{T} = -0.1$. The red curve represents the Cahn-Hilliard theory (phase separation) for $M=1$, $L=0$, $K=0$; the purple curve represents phase amplification for $M=1$, $L=1/127$, and $K=0$; the green curve represents the generalized Cahn-Hilliard theory in the presence of natural and forceful interconversion for $M=1$, $L=1/127$, $K=\num{1.3e-3}$. For the latter case (with forceful interconversion), the green circles indicate the three characteristic wavenumbers of the amplification factor: the maximum, $q_m$, the lower cut-off, $q_-$, and the upper cut-off, $q_+$. b) The time evolution of the structure factor, given by Eq.~(\ref{Eqn-SCH}), for the same parameters used in the generalized Cahn-Hilliard-Cook theory in the presence of natural and forceful interconversion. The black dotted line depicts the evolution of the maximum of the structure factor. Due to the external source of forceful interconversion, the maximum of the structure factor is interrupted at the wavenumber $q_-$, while for complete phase separation and phase amplification, the maximum of the structure factor will evolve to $q=0$ for an infinite-sized system.}
    \label{Fig_AmpFactor}
\end{figure}

The effect of interconversion on the temporal evolution of the concentration for a binary mixture of species A and B in the presence of interconversion is given in the simplest form \cite{MFT_PT_2021} as
\begin{equation}\label{Eqn_Gen_OrdParam_Evo}
    \pdv{\hat{c}_A}{t} = M\laplacian{\hat{\mu}}-L\hat{\mu} - K\hat{c}_A
\end{equation}
where $\hat{c}_A$ is the order parameter, related to the physical concentration of species A by $\hat{c}_A = 2c_A-1$, $M$ and $L$ are kinetic coefficients for the diffusion and natural (spontaneous) interconversion dynamics respectively, and the third term is an external source of forceful interconversion. We consider the case when interconversion, both natural and forceful, occurs through a reaction $A \ce{<=>} B$ where $K$ is the forward and reverse reaction rate. Physically, a source of forceful interconversion could be introduced via irradiation through photons \cite{Bellon_2001} which promote interconversion of species or it could be introduced through a flux of matter \cite{Verdasca_Chemically_1995}. In the absence of interconversion, when $L = 0$ and $K = 0$, then Eq. (\ref{Eqn_Gen_OrdParam_Evo}) reduces to the Cahn-Hilliard theory \cite{cahn_phase_1965}. Lastly, the reduced chemical potential difference, $\hat{\mu} = \hat{\mu}_A -\hat{\mu}_B$, is found from the variational derivative of the dimensionless Landau-Ginzburg free-energy functional \cite{MFT_PT_2021} as 
\begin{equation}\label{Eqn_Gen_Chem_Pot}
    \hat{\mu} = \frac{{\mu}}{k_\text{B}T_\text{c}} = \frac{1}{2}(1+\Delta \hat{T})\ln\left(\frac{1+\hat{c}_A}{1-\hat{c}_A}\right)-\hat{c}_A -\kappa\laplacian{\hat{c}_A}
\end{equation}
where the reduced distance to the critical temperature, $\Delta\hat{T}=1-T/T_\text{c}$, is negative in the spinodal (unstable) region, $k_B$ is Boltzmann's constant, and $\kappa$ is the square of the range of intermolecular interactions. In this work, we adopt $\kappa = 1$ in the units of the square of the molecular size. Expanding Eq.~(\ref{Eqn_Gen_Chem_Pot}) to first order and analytically evaluating Eq.~(\ref{Eqn_Gen_OrdParam_Evo}) via a Fourier transform, the characteristic growth rate, known as the ``amplification factor'' \cite{Shum_Phase_2021,MFT_PT_2021,cahn_phase_1965}, for spinodal decomposition in the presence of interconversion may be written in the form \cite{MFT_PT_2021}
\begin{equation}\label{Eqn-omega(q)}
    \omega(q,t) = M\kappa q_m^2(t) [q_m^2(t) -2q_-^2]-M\kappa [q^2-q_m^2(t)]^2
\end{equation}
where the two characteristic wavenumbers, $q_m$ and $q_-$, are the maximum and the lower cut-off of the amplification factor, respectively  (see Figure \ref{Fig_AmpFactor}a). Using a first order approximation, they have the form
\begin{equation}\label{Eqn_Char_Wavenums}
    q_m^2 = -\frac{(M\hat{\chi}_{q=0}^{-1}(t)+L\kappa)}{2M\kappa} \quad\text{and}\quad q_-^2 = -\frac{(K+L\Delta\hat{T})}{M\Delta\hat{T}+L\kappa}
\end{equation}
We note that the maximum of the amplification factor, $q_m  = q_m(t)$ is time dependent, while we hypothesize that $q_-$ is time independent. The time dependence of $q_m(t)$ is given through the higher order terms of the chemical potential, Eq.~(\ref{Eqn_Gen_Chem_Pot}), and is introduced into the time-dependent inverse susceptibility, $\hat{\chi}_{q=0}^{-1}(t) = \partial\hat{\mu}/\partial\hat{c}_A(t)$. The origin of this temporal evolution is due to the change in concentration at constant temperature from the unstable ($\hat{c}_A=0$) to the stable ($\hat{c}_A > 0$) regime; as such, in the second order approximation, $\hat{\chi}^{-1}_{q=0}(t) \simeq \Delta\hat{T} + (1+\Delta\hat{T})\hat{c}^2_A(t)$ \cite{MFT_PT_2021,Langer_Theory_1973,Billotet_Dynamic_1980}. In contrast, $q_-$ is an intrinsic property of the system, and since $q_-$ determines the cut-off for the smallest possible growing domain modes, then the steady-state limit of the time evolution of the maximum wavenumber will also be cut-off by $q_-$ as $q_m(t\to\infty) \propto q_-$. To verify this prediction, we numerically compute $q_m$ from the wavenumber corresponding to the maximum of the structure factor, $q_m^s$, and compare our results with the steady-state domain modes obtained from simulations of a nonequilibrium hybrid model.  

\subsection{Structure Factor Modified by Interconversion of Species}
Defining the order parameter fluctuation variable as $\delta \hat{c}(\vec{r},t) \equiv \hat{c}(\vec{r},t)-\hat{c}(\vec{r},t=0)$, the structure factor is given through the correlation function for the concentration fluctuations \cite{Bray_Theory_2002}; such that
\begin{equation}
	S(q,t) = \int \dd{\vec{r}}<\delta \hat{c}(\vec{r},t)\delta \hat{c}(\vec{r}_0,t)>e^{i\vec{q}\cdot\vec{r} }
\end{equation} 
As shown by Cook \cite{cook_brownian_1970} and Langer \textit{et al.} \cite{langer_new_1975}, the equation of motion for $S(q,t)$ is found by introducing order-parameter fluctuations into the time evolution of the order parameter, Eq. (\ref{Eqn_Gen_OrdParam_Evo}), with $\delta\hat{c}(\vec{r})$ as the fluctuation variable and spatially integrating $\langle|\delta\hat{c}|^2\rangle$. Following this procedure, we obtain the first-order solution for mixed diffusion-interconversion dynamics as
\begin{equation}
	\pdv{S(q,t)}{t} = 2\omega(q,t)S(q,t) + 2\left(Mq^2 + L\kappa\right)
\end{equation}
where $\omega(q,t)$ is given by Eq.~(\ref{Eqn-omega(q)}) \cite{Glotzer_consistent_1994,Coniglio_Multiscaling_1989,Coniglio_Novel_1990}. We note that in the absence of interconversion and forceful racemization, this equation reduces to the result presented by Cook \cite{cook_brownian_1970}. Solving this differential equation for the structure factor, assuming a linear approximation \cite{Langer_Theory_1973}, $\partial\omega/\partial t \ll \omega(q,t)$, gives \cite{Billotet_Dynamic_1980,Binder_Theory_1978,binder_collective_1983,Stobl_Structure_1985}
\begin{equation}\label{Eqn_Sqt_long}
    S(q,t) = S_\infty(q) - \left[S(q,t=0)+S_\infty(q)\right]e^{2\omega(q,t)t}
\end{equation}
where $S(q,t)$ represents the modified Cahn-Hilliard-Cook structure factor, which now includes natural forceful interconversion. In the limit of infinite time, when $\partial S(q,t)/\partial t=0$, the steady-state structure factor, $S_\infty(q)$, is given by
\begin{equation}\label{Eq-Schi}
	S(q,t\to\infty) = S_\infty(q) = \frac{Mq^2 + L\kappa}{-\omega(q,t\to\infty)}
\end{equation}
It can be seen that when either $L=0$ or $M=0$, then this equation in equilibrium conditions ($K=0$) reduces to the Ornstein-Zernike structure factor - $S_{OZ} =  \xi^2/(1+\xi^2q^2)$, where the correlation length of concentration fluctuations is $\xi^2 = -\kappa/\Delta\hat{T}$.

The time-dependent structure factor, Eq.~(\ref{Eqn_Sqt_long}), can be simplified by applying the condition that at $t=0$, the system is quenched from a sufficiently high temperature where $S(q,t=0)=0$. Therefore, Eq.~(\ref{Eqn_Sqt_long}) may be written as
\begin{equation}\label{Eqn-SCH}
	S(q,t) = S_{\infty}(q)\left(1 - e^{2\omega(q,t)t}\right) 
\end{equation}
which is valid from the initial stages of spinodal decomposition to the coarsening regime \cite{binder_collective_1983,Cahn_Later_1966}. Evaluating $\partial S_\infty/\partial q =0$ to determine the wavenumber corresponding to the maximum of the structure factor gives $q_m^s = 2^{1/4}q_-$ in the steady-state limit. The time evolution of the structure factor is illustrated in Figure~\ref{Fig_AmpFactor}b. To account for the time dependence of $q_m(t)$, we assume a simple approximation of the transition in the form $q_m(t)=q_m(t=0)\exp(-t/\tau)+q_-(1-\exp(-t/\tau))$ based on the limiting values of $q_m$ at $t=0$ and $t\to\infty$, where $\tau = 100$, is a system dependent parameter that controls the crossover from spinodal decomposition to the coarsening regime. As shown in Figure~\ref{Fig_AmpFactor}, the wavenumber corresponding to the maximum of the steady-state structure factor, $q_m^s$, aligns with the prediction of $q_-$ from the theory. To accurately match the predictions from the theory with the computational results presented in the following section, we scale the characteristic wavenumbers from the theory by the size of the system.

\begin{figure}[t]
    \centering
    \includegraphics[width=0.95\linewidth]{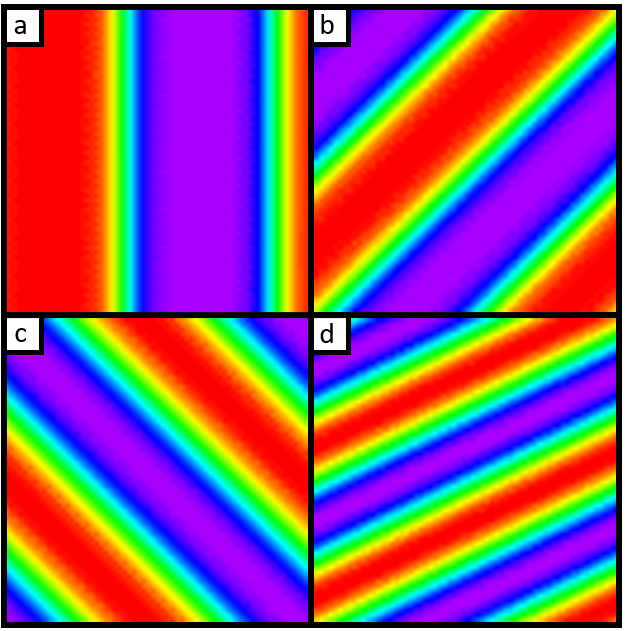}
    \caption{Steady-state phase domain morphology for different magnitudes of forceful interconversion (after $\sim 10^5$ time steps) computed from the time evolution of the order parameter, Eq.~(\ref{Eqn_Gen_OrdParam_Evo}), with $M=1$, $L=1/127$, $\Delta \hat{T} = -0.1$, $\ell=64$, $\sigma_i =0.1$, $\eta = 10^{-5}$. Morphologies are shown for the middle slice of the three-dimensional system at (a) $K=0$, (b) $K=\num{5e-4}$, (c) $K=\num{15e-4}$, and (d) $K=\num{25e-4}$. The red regions correspond to where the value of the normalized order parameter is $\hat{c}_A/\hat{c}_A^\text{max}=1$, the purple regions correspond to where the value of the normalized order parameter is $\hat{c}_A/\hat{c}_A^\text{min}=-1$, and the other colors depict the interface between these two regions. The image in (a) depicts a metastable structure toward phase amplification \cite{Shum_Phase_2021}, while the images in (b-d) are modulated steady-state structures with a characterize size, $1/q_-$.}
    \label{Fig_Morphology}
\end{figure}

\section{Methods}
\subsection{Spatial and Temporal Evolution of the Order Parameter}
Using the finite difference method \cite{Press_Numerical_2007}, with a spatial step $\Delta x = 1$ and a time step $\Delta t = 0.015$, we calculate the temporal evolution of the order parameter given by Eq.~(\ref{Eqn_Gen_OrdParam_Evo}) with a chemical potential given by Eq.~(\ref{Eqn_Gen_Chem_Pot}). We observed that for time steps $\Delta t > 0.015$, the solution diverges \cite{Press_Numerical_2007}. We include a random force term, $\eta$, to account for the thermal motion of the particles  \cite{cook_brownian_1970,binder_collective_1983}. The system is initialized on an $\ell\times\ell\times\ell$ cubic lattice with positions varied with initial random, Gaussian noise, $\sigma_i$. The structure factor is calculated via a Fast Fourier Transform (FFT) of the order parameter throughout the system \cite{Press_Numerical_2007}.

\subsection{Nonequilibrium Ising/Lattice-Gas Hybrid Model}
We consider an ``Ising-like'' lattice model where diffusion is modeled by ``swapping'' the position of two neighboring species and interconversion is modeled by ``flipping'' one species type to another \cite{Shum_Phase_2021}. The diffusion and interconversion dynamics are simulated using a hybrid of Kawasaki and Glauber Monte Carlo (MC) methods \cite{Bray_Theory_2002,kawasaki_diffusion_1966,glauber_timedependent_1963}, respectively. The species are arranged on an $\ell\times\ell\times\ell$ cubic lattice with a coordination number of 6. Using the Ising model Hamiltonian \cite{Huang_Stat_1987}
\begin{equation}
    H = -\frac{\varepsilon}{2}\sum_{i=1}^{\ell^3} \sum_{j\in \Omega(i)} s_i s_j
\end{equation}
where $s_i$, $s_j=\pm 1$ are spins, $\Omega(i)$ is the set of 6 nearest neighbors of spin $i$, and $\epsilon$ is the interaction energy. The critical temperature of this system is $T_\text{c} = 4.5115(1)\epsilon/k_\text{B}$ \cite{Heuer_Critical_1993}. Realizations are initialized with a random spin configuration in which $\ell^3/2$ are positive and the other $\ell^3/2$ are negative. In addition, we assume that at each MC step the probability of a random spin flip (a Glauber step) is $p_r$, while the probability of swapping a randomly selected pair of nearest neighbor spins (a Kawasaki step) is $1-p_r$. 

The equilibrium formulation, detailed in Ref. \cite{Shum_Phase_2021}, is converted to nonequilibrium via the introduction of an additional energy, $E$, incorporated into the Boltzmann factor for the probability that a spin flip will be accepted as $p\sim \exp\left[-(\Delta U-E)/k_\text{B}T\right]$, in which $\Delta U$ is the difference in internal energy of the system for this step \cite{metropolis_basic_1963}. Thus, the effect of the energy source only affects the interconversion dynamics of the system. The diffusion dynamics, determined in each Kawasaki step, occur with a probability that two spins will swap according to the Metropolis criterion without any additional energy source. We introduce a size-independent MC time as $t = n/\ell^3$, such that in every time unit, each spin in the system has a chance, $p_r$, to flip, or a chance, $1-p_r$, to swap with a neighboring spin. The probability of spin flipping is related to the diffusion and interconversion kinetic coefficients, $M$ and $L$ respectively, through $p_r = L/(M+L)$ \cite{Shum_Phase_2021}. Additionally, the frequency of spin flipping is absorbed into the time step, $\delta t$, so that the kinetic coefficients, and consequently $p_r$, do not depend on temperature. 

The dynamic structure factor is calculated using the method described in Kumar \textit{et al.} \cite{Kumar_Static_2005}. This method differed from the FFT method, used in Sec. 3.2, since the maximum of the structure factor (after normalizing by $\ell^3$) differed by a factor of $\pi$. Additionally, using the FFT method, the time evolution of the structure factor was interrupted at the wavenumber $q=1$ (indicating that the size of the two phase domains were half the size of the simulation box, $\ell/2$), while using the method presented in Kumar \textit{et al.}, the maximum of the structure factor was interrupted at a larger wavenumber. To correct for this difference, we scale the wavenumbers such that complete phase separation occurs at $q=1$.

\begin{figure}[t]
     \centering
    \includegraphics[width=0.49\linewidth]{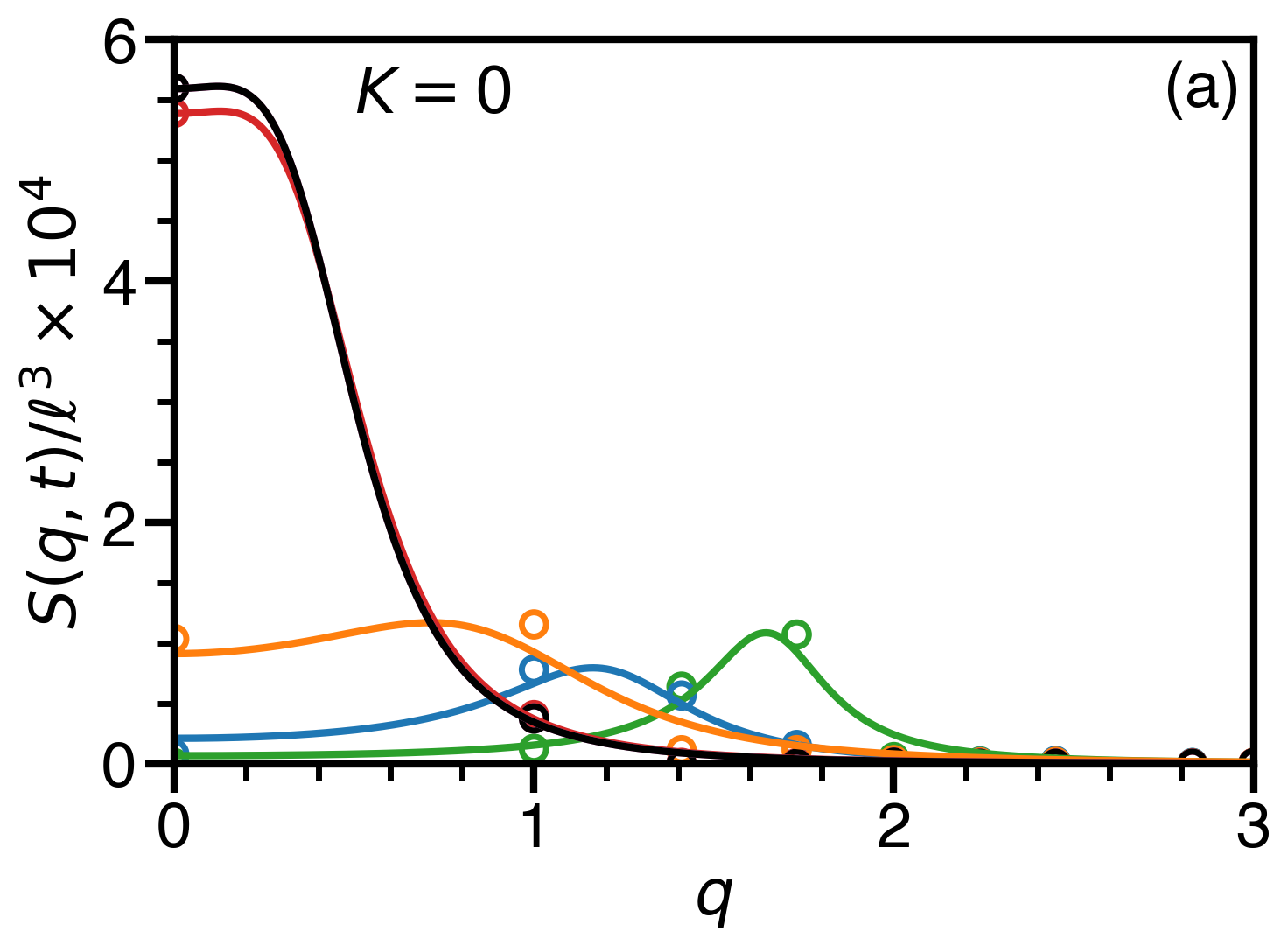}
    \includegraphics[width=0.49\linewidth]{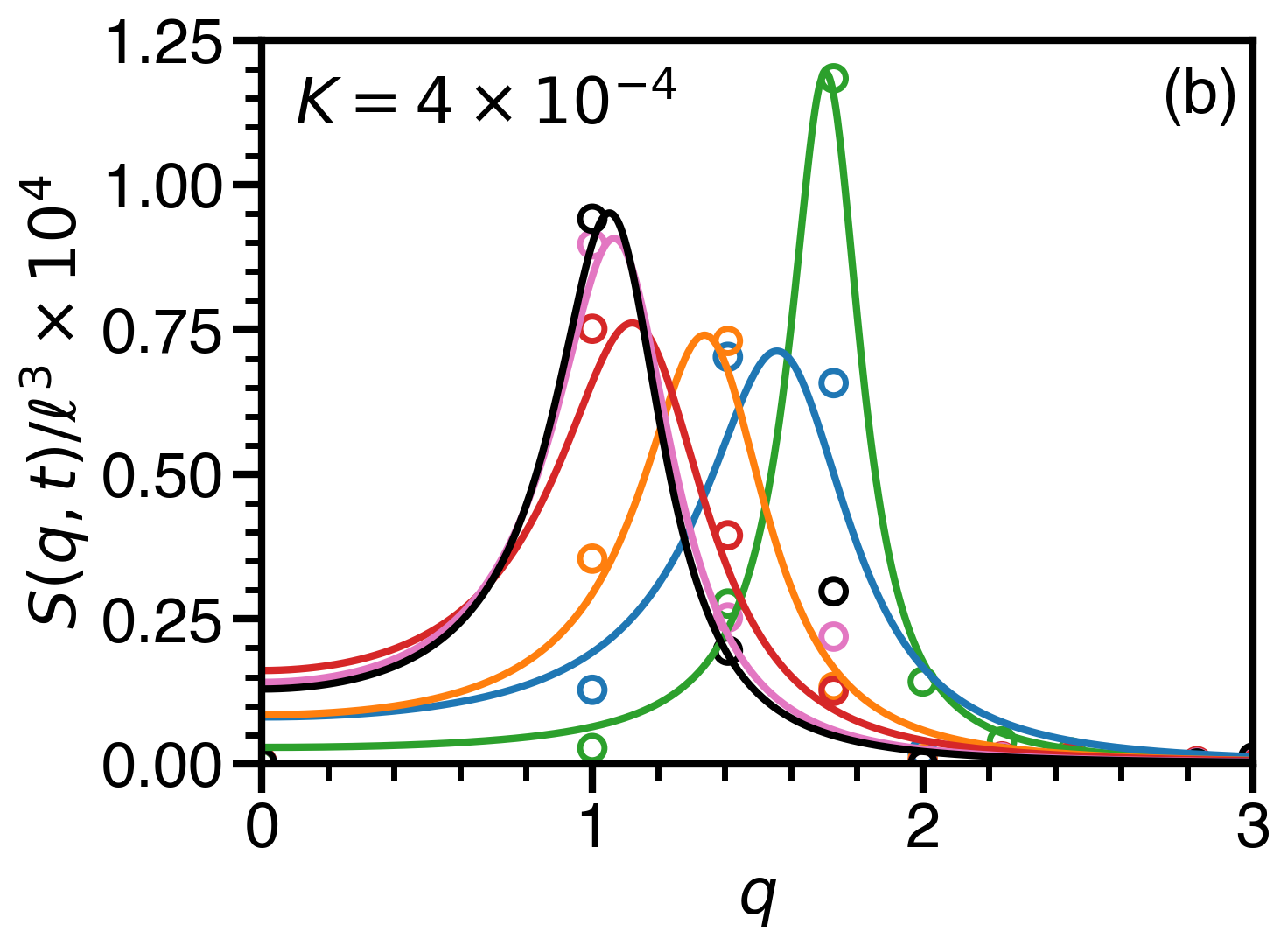}
    \includegraphics[width=0.49\linewidth]{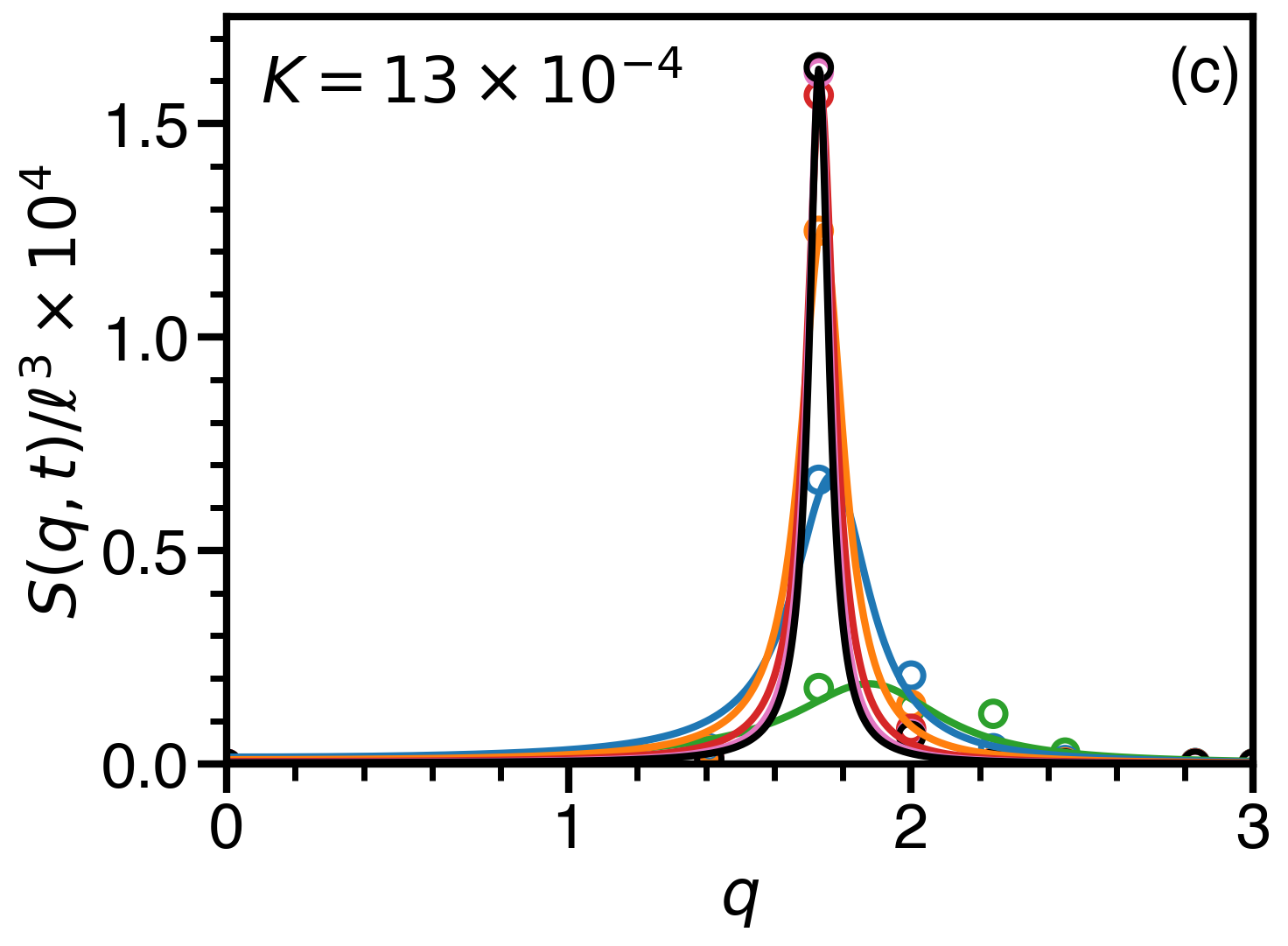}
    \includegraphics[width=0.49\linewidth]{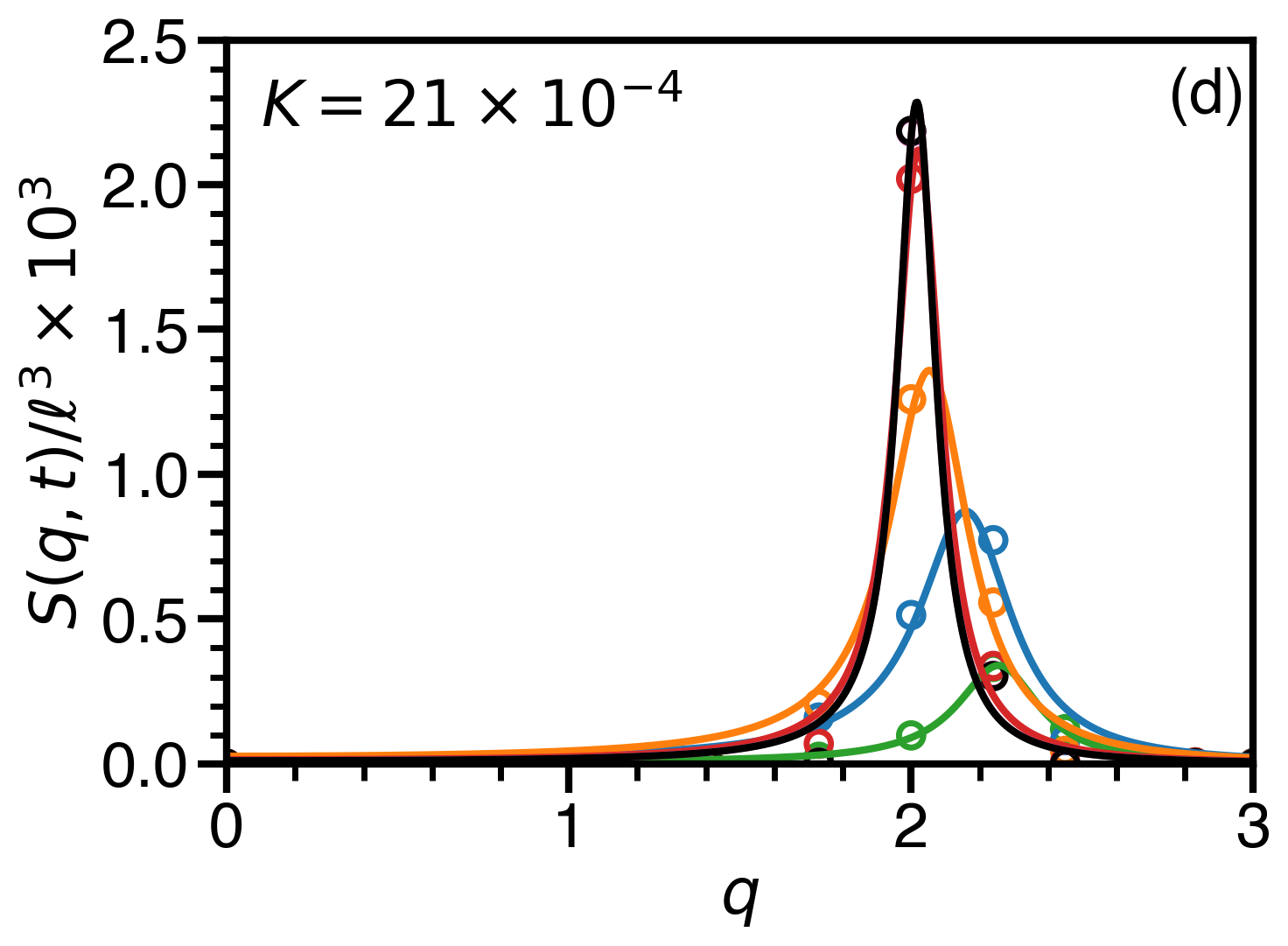}
    \caption{Time evolution of the structure factor computed from the Fast Fourier transform (FFT) of Eq.~(\ref{Eqn_Gen_OrdParam_Evo}) for $M=1$, $L=1/127$, $\Delta\hat{T}=-0.1$, $\ell=64$, $\sigma_i=0.1$, $\eta=10^{-5}$ depicted at times $t=\num{6e3}$ (green), $t=\num{1.2e4}$ (blue), $t=\num{2.4e4}$ (orange), $t=\num{5e4}$ (red), $t=\num{1e5}$ (pink), and $t=\num{2e5}$ (black). The open circles in (a-d) depict the computed structure factors for the four selected magnitudes of forceful interconversion averaged over $N=100$ realizations, while the solid lines illustrate the behavior of the structure factors assuming a Gaussian distribution. In (a), the evolution of the maximum of the structure factor is interrupted at the wavenumber $q=0$, corresponding to phase amplification, while in (b) the maximum is interrupted at $q=1$, corresponding to phase domains with a characteristic size of half the simulation box, $\ell/2$. In (c,d), the evolution of the maximum of the structure factor is interrupted at wavenumbers $q=q_->1$ corresponding to microphase separation.}
    \label{Fig_Sfact_Morphs}
\end{figure}

\section{Results and Discussion}
We confirmed that the presence of a source of forceful interconversion causes the system to phase separate into steady-state microphase domains as presented in Figures~\ref{Fig_Morphology}(a-d). Due to the periodic boundary conditions imposed in the continuum finite-difference method used to compute Eq.~(\ref{Eqn_Gen_OrdParam_Evo}), we found that the stripe morphologies will form at any angle with respect to the simulation box. The characteristic size of the stripe-like domains decreases with increasing forceful interconversion source strength, $K$. We note that a condition for microphase separation is that $K$ must be sufficiently ``strong'' as to overcome the natural interconversion. If the magnitude of $K$ is not strong enough, then (depending on the rate of natural interconversion) the system will either undergo phase amplification or complete phase separation. For instance, for $M=1$, $L=1/127$, $\Delta \hat{T} = -0.1$, and $K \le \num{4e-4}$, then microphase separation is not observed. Since $L=1/127$, the interconversion rate is relatively slow, and thus, the system has a higher probability of forming an interface between phases as shown in Figure~\ref{Fig_Morphology}a. However, for a system with natural interconversion this state is metastable, and  eventually, the interface between phases will break down and phase amplification will occur \cite{Shum_Phase_2021}. 

The time-dependent structure factors, which produce the stripe-like morphology, as illustrated in Figures~\ref{Fig_Morphology}(a-d), are presented in Figures~\ref{Fig_Sfact_Morphs}(a-d). We observe that the time evolution of the maximum of the structure factor in Figure~\ref{Fig_Sfact_Morphs}a is interrupted at the wavenumber $q = 0$, which corresponds to a system undergoing phase amplification. For $K=\num{4e-4}$ (Figure~\ref{Fig_Sfact_Morphs}b), the maximum of the time evolution of the structure factor is interrupted at $q=1$, indicating complete phase separation where the phase domains have a characteristic size of half the simulation box, $\ell/2$. In Figures~\ref{Fig_Sfact_Morphs}(c,d), the time evolution of the maximum of the structure factor is interrupted at higher wavenumbers depending on $K$. These wavenumbers correspond to the characteristic size of the stripe-like phase domains and are independent of the size of the system. The non-monotonic temporal evolution of the structure factor observed in Figures~\ref{Fig_Sfact_Morphs}b can be attributed to $\tau\approx \num{1.5e3}$, a large characteristic crossover time scale between spinodal decomposition and coarsening. This observation suggests that the crossover time scale, $\tau = \tau(K)$, may depend on forceful interconversion.

\begin{figure}[h!]
    \centering
    \includegraphics[width=\linewidth]{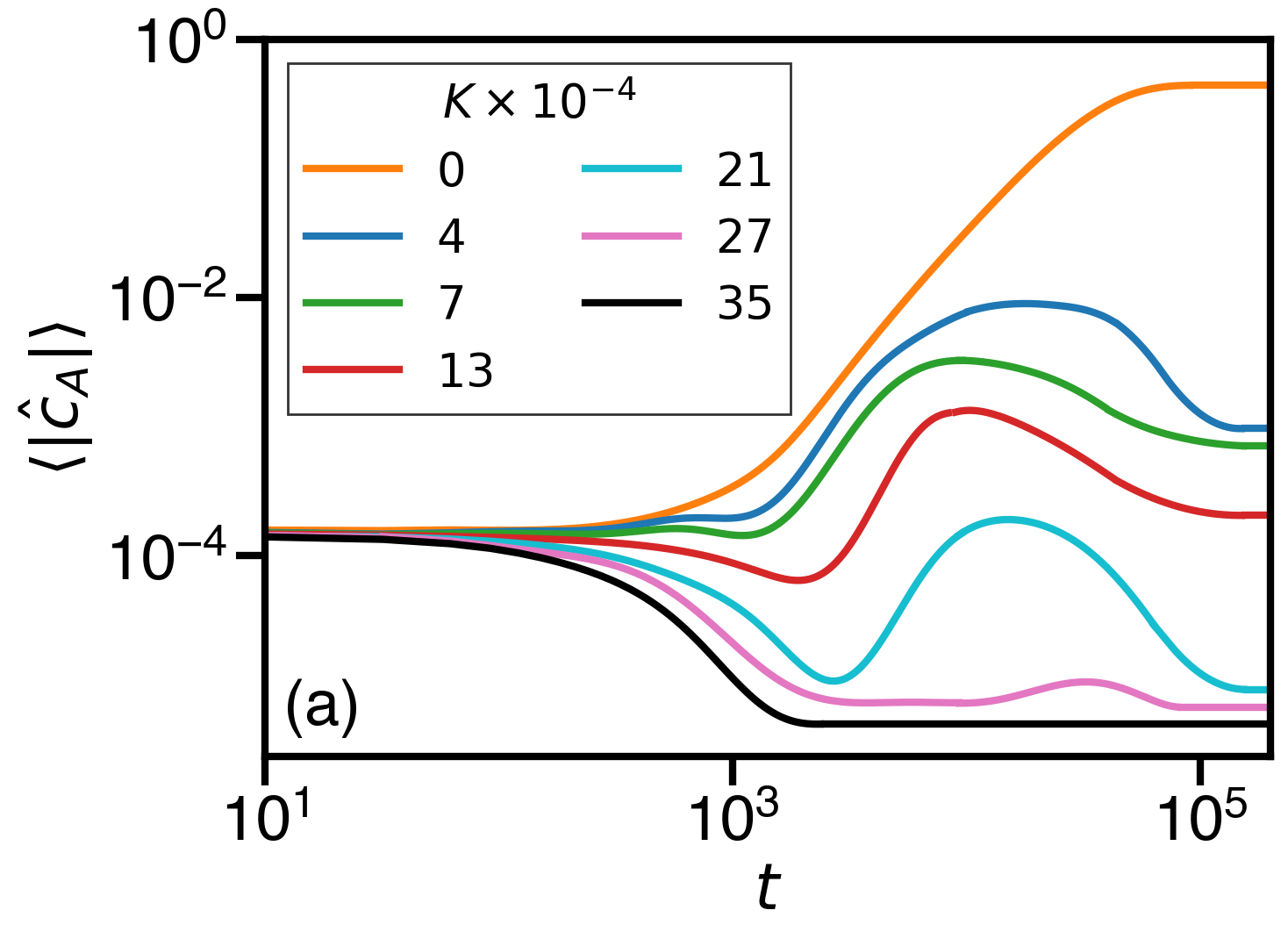}
    \includegraphics[width=\linewidth]{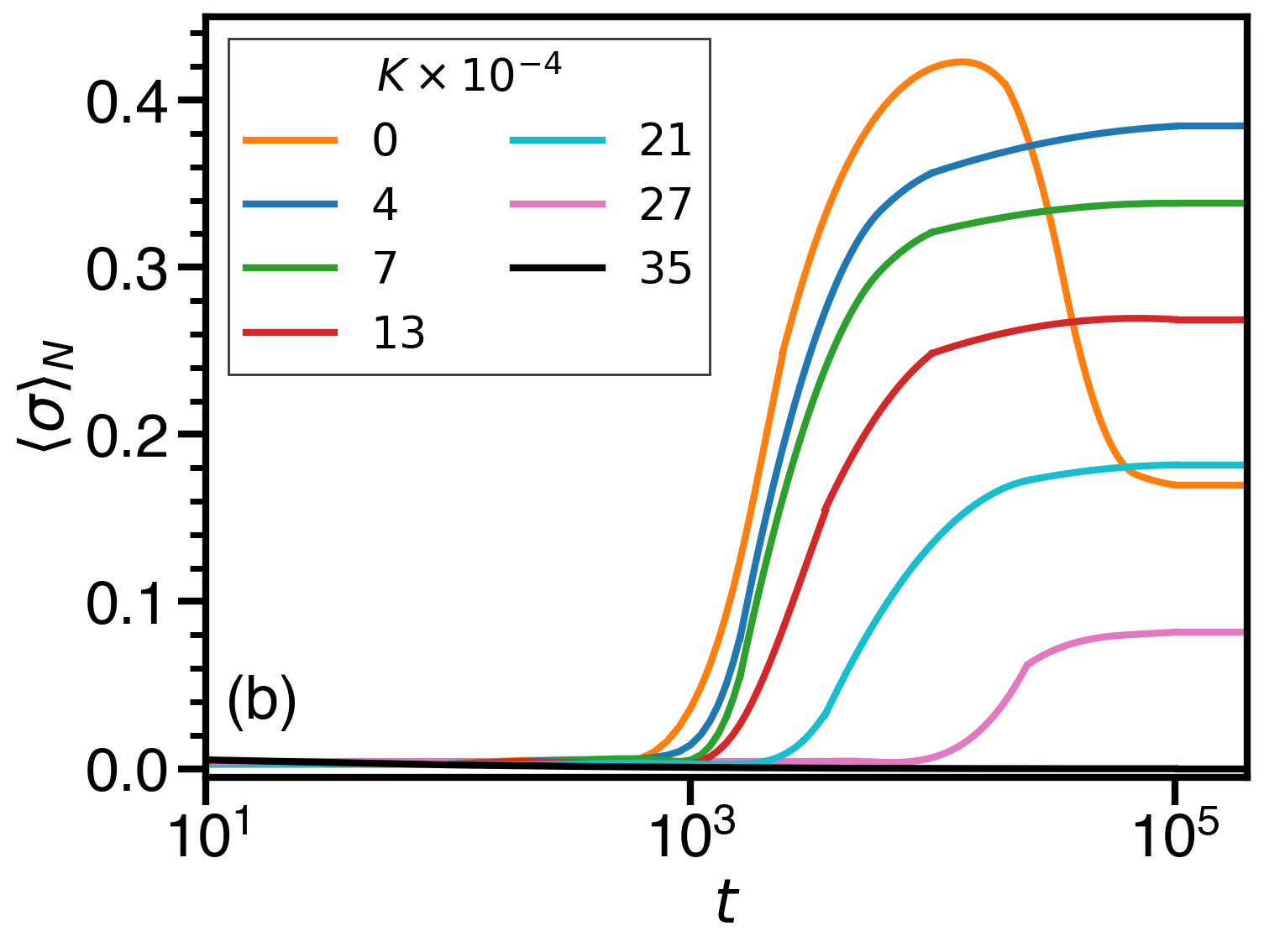}
    \caption{The temporal evolution of the symmetry of phase separation. a) The time evolution of the average order parameter, calculated by first averaging over all space and second by averaging the absolute value over $N=100$ realizations, for $M=1$, $L=1/127$, $\Delta T = -0.1$, $\sigma_i=0.1$, $\eta = \num{1.0e-5}$, and various magnitudes of forceful interconversion, $K$. b) The time evolution of the $N$-averaged standard deviation of the averaged order parameter, calculated by first determining the standard deviation of the spatially averaged order parameter and second by averaging over $N=100$ realizations. This method of averaging highlights the behavioral deviation from an equal concentration of species A and B, $\hat{c}_A=0$.}
    \label{Fig_AvgMag_Std}
\end{figure}

The temporal evolution of the order parameter was calculated from Eq.~(\ref{Eqn_Gen_OrdParam_Evo}) using the chemical potential given in Eq.~(\ref{Eqn_Gen_Chem_Pot}). The average value of the order parameter, calculated by first averaging over all space and second averaging the absolute value over $N=100$ realizations, is presented in Figure~\ref{Fig_AvgMag_Std}a. This method of averaging highlights the behavioral deviation from $\hat{c}_A=0$, when the concentration of species A is equivalent to species B; therefore, this figure represents the temporal evolution of the symmetry of phase separation. We observed that the initial value $\langle|\hat{c}_A|\rangle$ is determined by $\sigma_i$, the random initial configuration, whereas the steady-state behavior  of $\langle|{\hat{c}}_A|\rangle$ is determined by $\eta$, the thermal noise to be included in Eq.~(\ref{Eqn_Gen_OrdParam_Evo}). We find that $\langle|\hat{c}_A|\rangle$ develops a peak during the formation of the stripe-like patterns. As the phase domains coarsen, the averaged order parameter reaches a steady-state value, $\langle|\hat{c}_A(t\to\infty)|\rangle= c_0$ indicating the stable formation of the stripe-like domains. 

In Figure~\ref{Fig_AvgMag_Std}b, we show the temporal evolution of the standard deviation of the averaged order parameter, calculated by first determining the standard deviation over all space and second by averaging over $N=100$ realizations. We observed that the $N$-averaged standard deviation, $\langle\sigma\rangle_N$, was constant through the early stages of spinodal decomposition, but dramatically increased during the formation of the stripe-like patterns. We note that in the $K=0$ case due to phase amplification $\langle\sigma\rangle_N$ rapidly increases as the domains coarsen, but then decreases when one phase grows at the expense of the other. In this case, the constant steady-state limit of the averaged standard deviation, $\langle\sigma (t\to\infty)\rangle_N = \sigma_0$, indicates that the order parameter has reached it's equilibrium value, $|\hat{c}_A|=c_0$, which depends on the distance to the critical temperature.

\begin{figure}[h!]
    \centering
    \includegraphics[width=\linewidth]{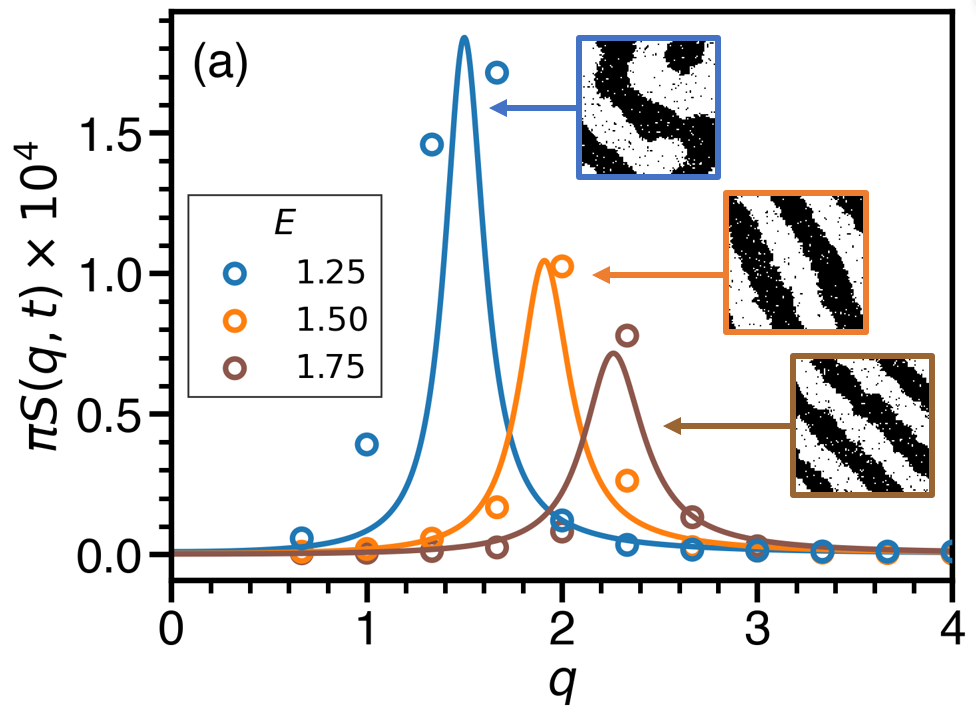}
    \includegraphics[width=\linewidth]{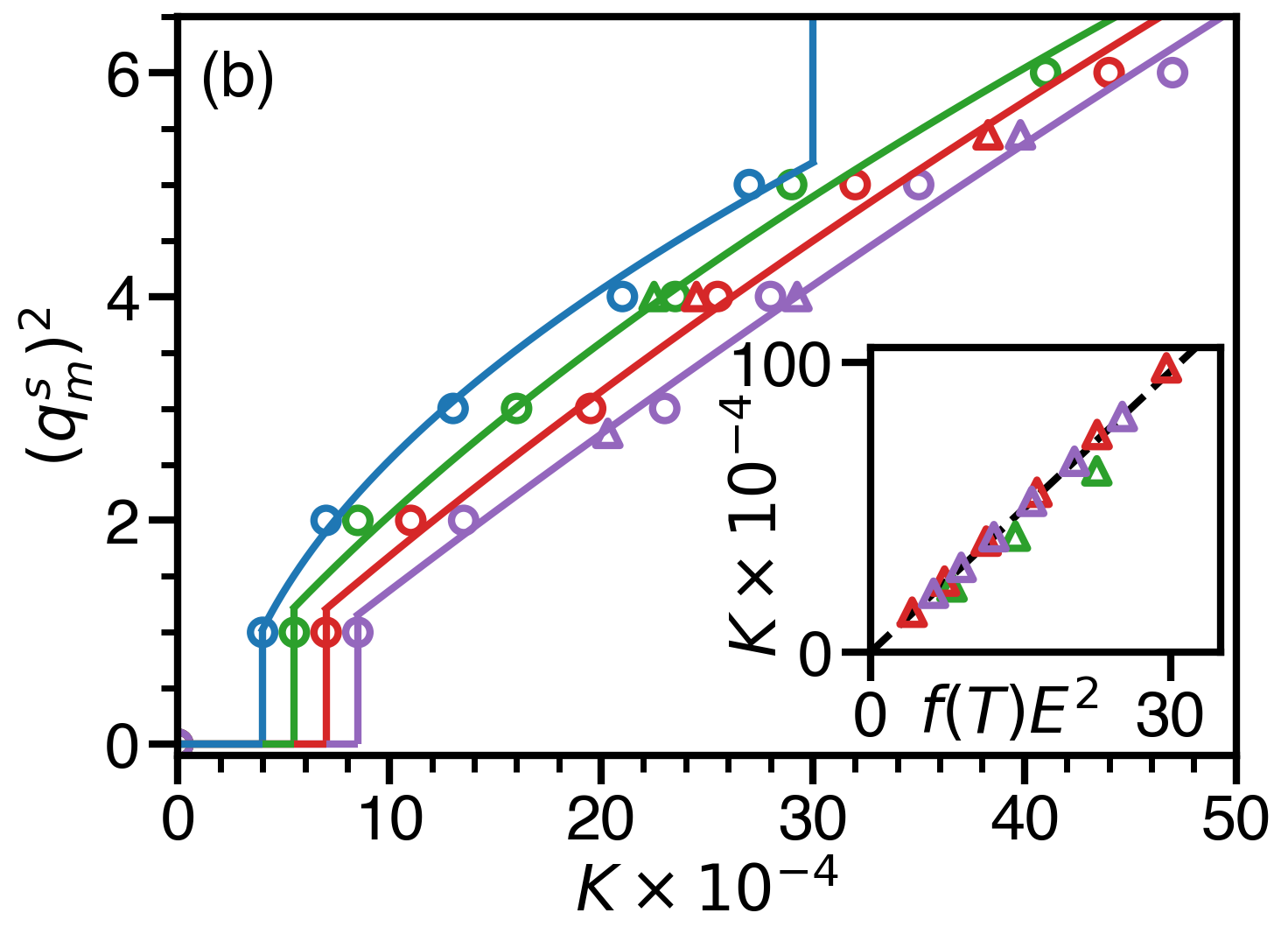}
    \caption{ a) Steady-state structure factors computed for the nonequilibrium hybrid model (open circles) and the prediction given by Eq.~(\ref{Eq_Empir_Schi}) (solid lines) for selected external energy sources ($E$) at $\Delta \hat{T} = -0.4$, $M=1$, $L=1/127$, $\ell = 100$, and averaged over $N=60$ realizations. The insets show steady-state ($t\sim \num{3e5}$) domain morphologies observed in the nonequilibrium hybrid model for the selected energies. b) The dependence of forceful interconversion on the wavenumber corresponding to the maximum of the structure factor, $q_m^s$, in the steady-state limit. The open circles are numerical computations of structure factors determined from FFTs of the time evolution of the order parameter, given by  Eq.~(\ref{Eqn_Gen_OrdParam_Evo}), in the steady-state limit ($t\sim 10^5$) for $M=1$, $L=1/127$, $\sigma_i =0.1$, and $\eta=10^{-5}$, averaged over $N=100$ realizations. The triangles correspond to the predictions of $K$ determined from fits of Eq.~(\ref{Eq_Empir_Schi}) to the structure factor for the nonequilibrium hybrid model, like those presented in (a). The curves illustrate the theoretical prediction $q_m(t\to\infty)\propto q_-$, given by the full expression for $q_-$, found from evaluating $\omega(q,0)=0$ using Eq.~(\ref{Eqn-omega(q)}). The inset shows the relationship between $K$ and $E$.}
    \label{Fig_Hybrid_Comp}
\end{figure}

In Figures~\ref{Fig_Hybrid_Comp}(a,b), we compare the structure factor theory with simulations of the nonequilibrium hybrid model (defined in Sec. 3.2). In Figure~\ref{Fig_Hybrid_Comp}a, we show the steady-state structure factor from simulations at three different additional energy values, $E$, at constant temperature, $\Delta\hat{T}=-0.4$, averaged over $N=60$ realizations. Unlike the morphologies computed from Eq.~(\ref{Eqn_Gen_OrdParam_Evo}) via the finite-difference method, the snapshots of the MC simulations shown in the insets of Figure~\ref{Fig_Hybrid_Comp}a depict stripe-like phase domains that form along the diagonal of the simulation box. We attribute this affect to the lattice structure utilized in the MC simulations. We introduce three system-dependent constants into the steady-state structure factor (when $q_m\sim q_-$) and use Eq.~(\ref{Eq-Schi}) in the form
\begin{equation}\label{Eq_Empir_Schi}
    S_\infty(q) = \frac{S_0a^2(q^2+L_\text{eff})}{a^2q_-^4+[q^2-(1+a)q_-^2]^2}
\end{equation}
where the amplitude ratio relating the theory to the nonequilibrium hybrid model is $S_0=46.5$ and the effective interconversion kinetic coefficient is $L_\text{eff}\sim L/M= 0.0012$. The constant $a = 0.2$, which describes the relationship between $q_m(t\to\infty)$ and $q_-$, broadens or sharpens the scattering peak. At the maximum of the structure factor, when $q = q_-$, then Eq.~(\ref{Eq_Empir_Schi}) reduces to $S_\infty(q_-) = S_0/(2q_-^2)$, which is independent of $a$. We note that in this form, Eq.~(\ref{Eq_Empir_Schi}) resembles the scattering intensity distribution of microemulsions \cite{Teubner_Origin_1987}. 

In Figure~\ref{Fig_Hybrid_Comp}b, the dependence of the wavenumber corresponding to the maximum of the structure factor, $q_m^s$, on the magnitude of forceful interconversion, $K$ is illustrated for the theoretical prediction (curves), computations of the time evolution of the order parameter (circles), and simulations of the nonequilibrium hybrid model (triangles). The curves are determined from the full expression for the lower cut-off wavenumber, $q_-$, found from Eq.~(\ref{Eqn-omega(q)}), when $\omega(q,0)=0$. A variable amplitude and shift are introduced to scale the theoretical prediction of $q_-$ such that microphase separation begins at $q=1$. An additional system dependent constant is introduced to describe the relationship between $q_m(t\to\infty)$ and $q_-$. The numerical computations of $q_m^s$ (averaged over $N=100$ realizations) from the time evolution of the order parameter, Eq.~(\ref{Eqn_Gen_OrdParam_Evo}), are shown in the steady-state regime (after $t\sim 10^5$ time steps). The magnitude of forceful interconversion, $K$, (for different external energies, $E$) was obtained for the nonequilibrium hybrid model using Eq.~(\ref{Eq_Empir_Schi}). In the insert of Figure~\ref{Fig_Hybrid_Comp}b, we illustrate the relationship between theory and the nonequilibrium hybrid model, as $K\propto E^2$.

In addition, as illustrated by Figure~\ref{Fig_Hybrid_Comp}b, we find two values of forceful interconversion, $K^*$ and $K^{**}$ (indicated by the vertical lines), that bound the formation of microphase domains. Both boundaries increase with $\Delta\hat{T}$; such that, for $\Delta \hat{T}>-0.1$, $K^{**}$ is located off the scale of the figure. For $K<K^*$ phase amplification was observed, while for $K>K^{**}$, no striped patterns were observed; instead, only a homogeneous solution persisted, in which an apparent structure on a small scale may be attributed to the correlations between concentration fluctuations. In this case, the particles are forced to interconvert so rapidly that diffusion is impossible. As shown in Figure~\ref{Fig_Hybrid_Comp}b, the lower bound, $K^*$, is associated with the characteristic wavenumber, $q^*=q=1$, which corresponds to phase domains that form at half the size of the simulation box, $\ell/2$. In contrast, the characteristic wavenumber associated with the upper bound, $q^{**}$, is strongly dependent on temperature \cite{Uralcan_Interconversion_2020,MFT_PT_2021}.

We note that for $\Delta\hat{T}=-0.1$, no structured microphase separation was observed in the nonequilibrium hybrid model. We attribute the lack of structured domains to the increase in concentration fluctuations facilitated by the close proximity to the critical point. This effect was not observed in the time evolution of the order parameter, shown in Figure~\ref{Fig_Morphology}(a-d), as the mean-field theory described in Sec. 2. is only applicable sufficiently far away from the critical point.

Lastly, as shown in Sec. 2, the wavenumber corresponding to the maximum of the structure factor, $q_m^s$, scales linearly with the lower cut-off wavenumber, $q_-$. Consequently, we observe the scaling law that $q_m^s\sim\sqrt{K}\propto \sqrt{f(T)} E$, where the temperature dependent prefactor is $f(T)\simeq 9.71 T/(T_\text{c}-T)$. This result, which verifies our initial hypothesis, has also been confirmed in studies of a chiral model where the source of forceful interconversion is established internally via dissipative intermolecular forces \cite{Uralcan_Interconversion_2020}. Interestingly, previous studies of phase separating block copolymers in the presence of forceful interconversion found that $q_m^s\sim K^{1/4}$ \cite{Christensen_Segregation_1996,Glotzer_consistent_1994}. As these previous studies considered an $n$-component order parameter to describe the block copolymer system (whereas, in this work, we describe our binary mixture via a single-component order parameter), this implies that the effect of $K$ on $q_m^s$ is system dependent and could depend on the nature of the order parameter.

\section{Conclusion}
We have demonstrated that the presence of a source of forceful interconversion on a binary system with diffusion and natural interconversion dynamics may produce microphase separation. We characterize the time evolution of the phase formation through two characteristic wavenumbers, $q_m$ and $q_-$, which correspond to the maximum and lower cut-off wavenumbers of the amplification factor from the generalized theory of spinodal decomposition. In the infinite time (steady-state) limit, we showed that $q_m(t\to\infty)\propto q_-\propto K^{1/2}$. We compared the structure factor theory with simulations of a nonequilibrium hybrid model that exhibits diffusion and natural interconversion dynamics and demonstrated that the source of forceful interconversion may be related to an external energy source,  as $K\propto E^2$, which allows interconversion to be more energetically favorable. The study of the domain formation in a binary system exhibiting a mixture of diffusion and interconversion dynamics could be used to study the dynamic behavior of polyamorphic states in supercooled water \cite{Gallo_Tale_2016,anisimov_thermodynamics_2018} as well as the vitreous states of various substances.

\section*{Acknowledgements}
The authors thank Bet\"ul Uralcan and Pablo Debenedetti for useful discussions. This work is a part of the research collaboration between the University of Maryland, Princeton University, Boston University, and Arizona State University supported by the National Science Foundation. The research at the University of Maryland was supported by NSF Award No. 1856479. The research at Boston University was supported by NSF Award No. 1856496. S.V.B. acknowledges the partial support of this research through the Dr. Bernard W. Gamson Computational Science Center at Yeshiva College.

\bibliography{references}

\begin{thebibliography}{10}
\expandafter\ifx\csname url\endcsname\relax
  \def\url#1{\texttt{#1}}\fi
\expandafter\ifx\csname urlprefix\endcsname\relax\def\urlprefix{URL }\fi
\expandafter\ifx\csname href\endcsname\relax
  \def\href#1#2{#2} \def\path#1{#1}\fi

\bibitem{Angell_TwoState_1971}
C.~A. Angell, Two-state thermodynamics and transport properties for water as
  zeroth-order results of a ``bond lattice'' model, J. Phys. Chem. 75~(24)
  (1971).
\newblock \href {https://doi.org/https://doi.org/10.1021/j100693a010}
  {\path{doi:https://doi.org/10.1021/j100693a010}}.

\bibitem{Angell_Amorphous_2004}
C.~A. Angell, Amorphous water, Annual Review of Physical Chemistry 55~(1)
  (2004) 559--583.
\newblock \href {https://doi.org/10.1146/annurev.physchem.55.091602.094156}
  {\path{doi:10.1146/annurev.physchem.55.091602.094156}}.

\bibitem{Gallo_Tale_2016}
P.~Gallo, K.~Amann-Winkel, C.~A. Angell, M.~A. Anisimov, F.~Caupin,
  C.~Chakravarty, E.~Lascaris, T.~Loerting, A.~Z. Panagiotopoulos, J.~Russo,
  J.~A. Sellberg, H.~E. Stanley, H.~Tanaka, C.~Vega, L.~Xu, L.~G.~M.
  Pettersson, Water: A tale of two liquids, Chem. Rev. 116~(13) (2016)
  7463--7500.

\bibitem{Sastry_Liquid_2003}
S.~Sastry, C.~A. Angell, Liquid–liquid phase transition in supercooled
  silicon, Nat. Mater. 2 (2003) 739--743.
\newblock \href {https://doi.org/10.1038/nmat994} {\path{doi:10.1038/nmat994}}.

\bibitem{Xu_Thermodynamics_2006}
L.~Xu, S.~V. Buldyrev, C.~A. Angell, H.~E. Stanley,
  \href{https://link.aps.org/doi/10.1103/PhysRevE.74.031108}{Thermodynamics and
  dynamics of the two-scale spherically symmetric jagla ramp model of anomalous
  liquids}, Phys. Rev. E 74 (2006) 031108.
\newblock \href {https://doi.org/10.1103/PhysRevE.74.031108}
  {\path{doi:10.1103/PhysRevE.74.031108}}.
\newline\urlprefix\url{https://link.aps.org/doi/10.1103/PhysRevE.74.031108}

\bibitem{Bhat_Vitrification_2007}
M.~H. Bhat, V.~Molinero, E.~Soignard, V.~C. Solomon, S.~Sastry, J.~L. Yarger,
  C.~A. Angell, \href{https://doi.org/10.1038/nature06044}{Vitrification of a
  monatomic metallic liquid}, Nature 448 (2007) 787--790.
\newline\urlprefix\url{https://doi.org/10.1038/nature06044}

\bibitem{Xu_Monatomic_2009}
L.~Xu, S.~V. Buldyrev, N.~Giovambattista, C.~A. Angell, H.~E. Stanley, A
  monatomic system with a liquid-liquid critical point and two distinct glassy
  states, J. Chem. Phys. 130 (2009) 054505.
\newblock \href {https://doi.org/https://doi.org/10.1063/1.3043665}
  {\path{doi:https://doi.org/10.1063/1.3043665}}.

\bibitem{Lascaris_Search_2014}
E.~Lascaris, M.~Hemmati, S.~V. Buldyrev, H.~E. Stanley, C.~A. Angell,
  \href{https://doi.org/10.1063/1.4879057}{Search for a liquid-liquid critical
  point in models of silica}, J. Chem. Phys. 140 (2014) 224502.
\newline\urlprefix\url{https://doi.org/10.1063/1.4879057}

\bibitem{Lascaris_Diffusivity_2015}
E.~Lascaris, M.~Hemmati, S.~V. Buldyrev, H.~E. Stanley, C.~A. Angell,
  \href{https://doi.org/10.1063/1.4913747}{Diffusivity and short-time dynamics
  in two models of silica}, J. Chem. Phys. 142 (2015) 104506.
\newline\urlprefix\url{https://doi.org/10.1063/1.4913747}

\bibitem{Pool_Polyamorphic_1997}
P.~H. Poole, T.~Grande, C.~A. Angell, P.~F. McMillan, Polymorphic phase
  transitions in liquids and glasses, Science 275 (1997) 322--323.
\newblock \href {https://doi.org/10.1126/science.275.5298.322}
  {\path{doi:10.1126/science.275.5298.322}}.

\bibitem{Tanaka_Liquid_2020}
H.~Tanaka, Liquid–liquid transition and polyamorphism, J. Chem. Phys. 153
  (2020) 130901.
\newblock \href {https://doi.org/https://doi.org/10.1063/5.0021045}
  {\path{doi:https://doi.org/10.1063/5.0021045}}.

\bibitem{anisimov_thermodynamics_2018}
M.~A. Anisimov, M.~Duska, F.~Caupin, L.~E. Amrhein, A.~Rosenbaum, R.~J. Sadus,
  Thermodynamics of {Fluid} {Polyamorphism}, Physical Review X 8~(1) (2018)
  011004.
\newblock \href {https://doi.org/10.1103/PhysRevX.8.011004}
  {\path{doi:10.1103/PhysRevX.8.011004}}.

\bibitem{Shum_Phase_2021}
N.~A. Shumovskyi, T.~J. Longo, S.~V. Buldyrev, M.~A. Anisimov, Phase
  amplification in spinodal decomposition of immiscible fluids with
  interconversion of species, Phys. Rev. E 103 (2021) L060101.
\newblock \href {https://doi.org/10.1103/PhysRevE.103.L060101}
  {\path{doi:10.1103/PhysRevE.103.L060101}}.

\bibitem{Petsev_Effect_2021}
N.~D. Petsev, F.~H. Stillinger, P.~G. Debenedetti1, Effect of
  configuration-dependent multi-body forces on interconversion kinetics of a
  chiral tetramer model, J. Chem. Phys. 155 (2021) 084105.
\newblock \href {https://doi.org/https://doi.org/10.1063/5.0060266}
  {\path{doi:https://doi.org/10.1063/5.0060266}}.

\bibitem{Uralcan_Interconversion_2020}
B.~Uralcan, T.~J. Longo, M.~A. Anisimov, F.~H. Stillinger, P.~G. Debenedetti,
  \href{https://arxiv.org/abs/2109.06925}{Interconversion-controlled
  liquid-liquid phase separation in a molecular chiral model}, ArXiv (2021).
\newline\urlprefix\url{https://arxiv.org/abs/2109.06925}

\bibitem{MFT_PT_2021}
T.~J. Longo, M.~A. Anisimov, \href{https://arxiv.org/abs/2102.11148}{Phase
  transitions affected by molecular interconversion}, ArXiv (July 2021).
\newblock \href {https://doi.org/2102.11148v4} {\path{doi:2102.11148v4}}.
\newline\urlprefix\url{https://arxiv.org/abs/2102.11148}

\bibitem{glotzer_monte_1994}
S.~C. Glotzer, D.~Stauffer, N.~Jan, Monte {Carlo} simulations of phase
  separation in chemically reactive binary mixtures, Physical Review Letters
  72~(26) (1994) 4109--4112.
\newblock \href {https://doi.org/10.1103/PhysRevLett.72.4109}
  {\path{doi:10.1103/PhysRevLett.72.4109}}.

\bibitem{glotzer_reaction-controlled_1995}
S.~C. Glotzer, E.~A. Di~Marzio, M.~Muthukumar, Reaction-{Controlled}
  {Morphology} of {Phase}-{Separating} {Mixtures}, Physical Review Letters
  74~(11) (1995) 2034--2037.
\newblock \href {https://doi.org/10.1103/PhysRevLett.74.2034}
  {\path{doi:10.1103/PhysRevLett.74.2034}}.

\bibitem{Christensen_Segregation_1996}
J.~J. Christensen, K.~Elder, H.~C. Fogedby,
  \href{https://link.aps.org/doi/10.1103/PhysRevE.54.R2212}{Phase segregation
  dynamics of a chemically reactive binary mixture}, Phys. Rev. E 54 (1996)
  R2212--R2215.
\newblock \href {https://doi.org/10.1103/PhysRevE.54.R2212}
  {\path{doi:10.1103/PhysRevE.54.R2212}}.
\newline\urlprefix\url{https://link.aps.org/doi/10.1103/PhysRevE.54.R2212}

\bibitem{cahn_phase_1965}
J.~W. Cahn, Phase {Separation} by {Spinodal} {Decomposition} in {Isotropic}
  {Systems}, The Journal of Chemical Physics 42~(1) (1965) 93--99.
\newblock \href {https://doi.org/10.1063/1.1695731}
  {\path{doi:10.1063/1.1695731}}.

\bibitem{Bellon_2001}
R.~A. Enrique, P.~Bellon, Compositional patterning in immiscible alloys driven
  by irradiation, Physical Review B 63~(134111) (2001) 45--65.
\newblock \href {https://doi.org/10.1103/PhysRevB.63.134111}
  {\path{doi:10.1103/PhysRevB.63.134111}}.

\bibitem{Verdasca_Chemically_1995}
J.~Verdasca, P.~Borckmans, G.~Dewel,
  \href{https://link.aps.org/doi/10.1103/PhysRevE.52.R4616}{Chemically frozen
  phase separation in an adsorbed layer}, Phys. Rev. E 52 (1995) R4616--R4619.
\newblock \href {https://doi.org/10.1103/PhysRevE.52.R4616}
  {\path{doi:10.1103/PhysRevE.52.R4616}}.
\newline\urlprefix\url{https://link.aps.org/doi/10.1103/PhysRevE.52.R4616}

\bibitem{Langer_Theory_1973}
J.~Langer, M.~Bar-on, Theory of early-stage spinodal decomposition, Annals of
  Physics 78~(2) (1973) 421--452.
\newblock \href {https://doi.org/https://doi.org/10.1016/0003-4916(73)90266-2}
  {\path{doi:https://doi.org/10.1016/0003-4916(73)90266-2}}.

\bibitem{Billotet_Dynamic_1980}
C.~Billotet, K.~Binder, Dynamic correlation of fluctuations during spinodal
  decomposition, Phys. A: Stat. Mech. Appl. 103~(1) (1980) 99--118.
\newblock \href {https://doi.org/https://doi.org/10.1016/0378-4371(80)90209-5}
  {\path{doi:https://doi.org/10.1016/0378-4371(80)90209-5}}.

\bibitem{Bray_Theory_2002}
A.~J. Bray, Theory of phase-ordering kinetics, Adv. Phys. 51~(2) (2002)
  481--587.
\newblock \href {https://doi.org/10.1080/00018730110117433}
  {\path{doi:10.1080/00018730110117433}}.

\bibitem{cook_brownian_1970}
H.~E. Cook, Brownian motion in spinodal decomposition, Acta Metallurgica 18~(3)
  (1970) 297--306.
\newblock \href {https://doi.org/10.1016/0001-6160(70)90144-6}
  {\path{doi:10.1016/0001-6160(70)90144-6}}.

\bibitem{langer_new_1975}
J.~S. Langer, M.~Bar-on, H.~D. Miller, New computational method in the theory
  of spinodal decomposition, Physical Review A 11~(4) (1975) 1417--1429.
\newblock \href {https://doi.org/10.1103/PhysRevA.11.1417}
  {\path{doi:10.1103/PhysRevA.11.1417}}.

\bibitem{Glotzer_consistent_1994}
S.~C. Glotzer, A.~Coniglio,
  \href{https://link.aps.org/doi/10.1103/PhysRevE.50.4241}{Self-consistent
  solution of phase separation with competing interactions}, Phys. Rev. E 50
  (1994) 4241--4244.
\newblock \href {https://doi.org/10.1103/PhysRevE.50.4241}
  {\path{doi:10.1103/PhysRevE.50.4241}}.
\newline\urlprefix\url{https://link.aps.org/doi/10.1103/PhysRevE.50.4241}

\bibitem{Coniglio_Multiscaling_1989}
A.~Coniglio, M.~Zannetti,
  \href{https://doi.org/10.1209/0295-5075/10/6/012}{Multiscaling in growth
  kinetics}, Europhysics Letters ({EPL}) 10~(6) (1989) 575--580.
\newblock \href {https://doi.org/10.1209/0295-5075/10/6/012}
  {\path{doi:10.1209/0295-5075/10/6/012}}.
\newline\urlprefix\url{https://doi.org/10.1209/0295-5075/10/6/012}

\bibitem{Coniglio_Novel_1990}
A.~Coniglio, M.~Zannetti,
  \href{https://www.sciencedirect.com/science/article/pii/037843719090341O}{Novel
  dynamical scaling in kinetic growth phenomena}, Physica A: Statistical
  Mechanics and its Applications 163~(1) (1990) 325--333.
\newblock \href {https://doi.org/https://doi.org/10.1016/0378-4371(90)90341-O}
  {\path{doi:https://doi.org/10.1016/0378-4371(90)90341-O}}.
\newline\urlprefix\url{https://www.sciencedirect.com/science/article/pii/037843719090341O}

\bibitem{Binder_Theory_1978}
K.~Binder, C.~Billotet, P.~Mirold, On the theory of spinodal decomposition in
  solid and liquid binary mixtures, Zeitschrift für Physik B Condensed Matter
  30~(2) (1978) 183--195.

\bibitem{binder_collective_1983}
K.~Binder, Collective diffusion, nucleation, and spinodal decomposition in
  polymer mixtures, J. Chem. Phys. 79 (1983) 6387.
\newblock \href {https://doi.org/10.1063/1.445747}
  {\path{doi:10.1063/1.445747}}.

\bibitem{Stobl_Structure_1985}
G.~R. Strobl, Structure evolution during spinodal decomposition of polymer
  blends, Macromol. 18 (1985) 559--563.

\bibitem{Cahn_Later_1966}
J.~W. Cahn, The later stages of spinodal decomposition and the beginnings of
  particle coarsening, Acta Metallurgica 14 (1966) 1685--1962.

\bibitem{Press_Numerical_2007}
W.~H. Press, S.~A. Teukolsky, W.~T. Vetterling, B.~P. Flannery, Numerical
  Recipes: The Art of Scientific Computing, 3rd Edition, Cambridge Univesity
  Press, Cambridge, United Kingdom, 2007.

\bibitem{kawasaki_diffusion_1966}
K.~Kawasaki, \href{https://link.aps.org/doi/10.1103/PhysRev.145.224}{Diffusion
  {Constants} near the {Critical} {Point} for {Time}-{Dependent} {Ising}
  {Models}. {I}}, Physical Review 145~(1) (1966) 224--230.
\newblock \href {https://doi.org/10.1103/PhysRev.145.224}
  {\path{doi:10.1103/PhysRev.145.224}}.
\newline\urlprefix\url{https://link.aps.org/doi/10.1103/PhysRev.145.224}

\bibitem{glauber_timedependent_1963}
R.~J. Glauber,
  \href{http://aip.scitation.org/doi/abs/10.1063/1.1703954}{Time‐{Dependent}
  {Statistics} of the {Ising} {Model}}, Journal of Mathematical Physics 4~(2)
  (1963) 294--307.
\newblock \href {https://doi.org/10.1063/1.1703954}
  {\path{doi:10.1063/1.1703954}}.
\newline\urlprefix\url{http://aip.scitation.org/doi/abs/10.1063/1.1703954}

\bibitem{Huang_Stat_1987}
K.~Huang, Statistical Mechanics, 2nd Edition, John Wiley and Sons, Inc., 1987.

\bibitem{Heuer_Critical_1993}
H.-O. Heuer, Critical crossover phenomena in disordered ising systems, J. Phys.
  A: Math. Gen. 26~(6) (1993) L333--L339.
\newblock \href {https://doi.org/10.1088/0305-4470/26/6/007}
  {\path{doi:10.1088/0305-4470/26/6/007}}.

\bibitem{metropolis_basic_1963}
N.~Metropolis, R.~L. Ashenhurst, Basic {Operations} in an {Unnormalized}
  {Arithmetic} {System}, IEEE Transactions on Electronic Computers EC-12~(6)
  (1963) 896--904.
\newblock \href {https://doi.org/10.1109/PGEC.1963.263593}
  {\path{doi:10.1109/PGEC.1963.263593}}.

\bibitem{Kumar_Static_2005}
P.~Kumar, S.~V. Buldyrev, F.~Sciortino, E.~Zaccarelli, H.~E. Stanley,
  \href{https://link.aps.org/doi/10.1103/PhysRevE.72.021501}{Static and dynamic
  anomalies in a repulsive spherical ramp liquid: Theory and simulation}, Phys.
  Rev. E 72 (2005) 021501.
\newblock \href {https://doi.org/10.1103/PhysRevE.72.021501}
  {\path{doi:10.1103/PhysRevE.72.021501}}.
\newline\urlprefix\url{https://link.aps.org/doi/10.1103/PhysRevE.72.021501}

\bibitem{Teubner_Origin_1987}
M.~Teubner, R.~Strey, Origin of the scattering peak in microemulsions, J. Chem.
  Phys. 87 (1987) 3195.
\newblock \href {https://doi.org/https://doi.org/10.1063/1.453006}
  {\path{doi:https://doi.org/10.1063/1.453006}}.

\end{thebibliography}


\end{document}